\documentclass{article}

\usepackage[]{fullpage}

\usepackage{color, graphicx, url, amssymb}

\begin{document}

\title{Multi-agent simulation of voter’s behaviour}

\author{Albin Soutif \qquad Carole Adam \qquad Sylvain Bouveret \\ Univ. Grenoble-Alpes, Grenoble Informatics Laboratory}

\date{\emph{\small{This is an internship report originally written in June 2018 by A. Soutif, ENSIMAG intern}}\\
\emph{\small{under the supervision of C. Adam and S. Bouveret}}
}

\maketitle


\begin{abstract}
    The goal of this paper is to simulate the voters’ behaviour given a voting method. Our approach uses a multi-agent simulation in order to model a voting process through many iterations, so that the voters can vote by taking into account the results of polls. Here we only tried basic rules and a single voting method, but further attempts could explore new features.
    
    \textbf{Keywords:} Computational social choice, Iterative voting, multi-agent simulation
\end{abstract}

\section{INTRODUCTION}

A voting process involves the participation of many people that interact together in order to reach a common decision. In this paper, we focus on voting processes in which a single person is elected. A voting method is defined as the set of rules that determine the winner of the election, given an input from each voter, for example their preferred candidate or an order relation between all candidates. 
Social Choice Theory is the field that studies the aggregation of individual preferences towards a collective choice, like for example electing a candidate or choosing a movie. Computational social choice is a recent field which aim is to apply computer science to social choice problems \cite{BCE16}. Here we aim to design a multi-agent simulation in order to study the collective decision process involved in an election.

A multi-agent simulation is a simulation in which we make several autonomous agent interact with each other in a defined environment. The goal of this kind of simulation is to study the global behaviour emerging from the local actions of the agents \cite{ea96,mn10,macal16}.
There are only a few attempts to mix the field of multi-agent simulation and computational social choice to model a voting process, and that is the kind of approach we use in this paper. The goal of this approach is to study the dynamics of voting methods through many iterations. Here, we will focus on one particular voting method which is plurality voting. In this method, voters vote for only one candidate and the winner of the election is the candidate with the most votes.

Our approach was supported by the results of a study on the French presidential election of 2017 \cite{vot2017}. In this study, people were asked to vote with different voting methods, and to tell for whom they officially voted. The goal of this study was to better understand the properties of several voting methods. Among these voting methods, one in particular helped to build this model, because it gave a precise insight of voters’ voting intentions (not just their favourite candidate), it is the continuous evaluation method. In this method, a voter gives a mark between 0 and 100 to each of the 11 candidates. Thanks to this study, we had access to over 30 000 of these continuous evaluations, which helped us initialise our model.

A voting process can be modelled intuitively as a multi-agent system. Indeed, voters and candidates can be modelled as autonomous agents, the real world, or the internet, is the environment in which these agents evolve, and the political debates, polls, conversations ... Are the interactions which can influence their final decision. In a multi-agent simulation, it is reasonable to limit ourselves to a simpler model, in which the agents evolve in a limited environment and have much simpler interactions with each other. Otherwise it would be too difficult to model, to interpret, and too computationally expensive.

First we will review the state of the art in this field (Section~\ref{sec:soa}), then we will present the way we built our model (Section~\ref{sec:mod}), then our results along with our interpretation (Section~\ref{sec:rez}), and eventually we will conclude and discuss some possible extensions of the model (Section~\ref{sec:cci}).

\section{STATE OF THE ART} \label{sec:soa}

Some interesting attempts have been made at building voting process models. Here we will review the ones that inspired us the most in building our model.

\subsection{Static models}

A static simulation has been made by Ka-Ping Yee \cite{yee} in which the voters are modelled by 2-dimensional points. In this simulation a normal distribution of voters is drawn around every point, and each point is coloured according to the candidate that won the election with that distribution of voters. The conclusion of these simulations highlighted the different behaviours of some voting rules.

\subsection{Iterative models}

Airiau et al. \cite{agp17} attempted to model the outcome of an election where voters could react strategically by learning from the others and trying to react in order to maximise an utility function. The authors found out that for a given set of rule and a precise methodology, the outcome was convergent and had good mathematical properties. The drawback is that for the convergence to be certain, the voters had to vote one by one and know at each step which candidate was winning, which is not a realistic procedure.

Another attempt has been made for educational purpose by a network of researchers in multi-agent simulation (MAPS) \cite{rm2014}. They made a multi-agent simulation called ``the Iznogoud model'', with simple interactions between 2-dimensional agents. The interactions used were local (each voter directly influenced its surroundings), whereas in our model there are no local interactions because we only focused on the global effect of polls, and not on the effect of local interactions such as political conversations between individuals. Also in the Iznogoud model, the voters were initialised randomly, making it impossible to compare the results to any real case scenario.

In both the approaches from Ka-Ping Yee \cite{yee} and the Iznogoud simulation \cite{rm2014}, the voters are placed in a 2-dimensional space. In Ka-Ping Yee simulation the model is not dynamic, so such a model would not help us study the dynamics of voting methods. In the Iznogoud simulation, the interactions between voters are local and very simple, and the model is initialised with a random population of voters. In our model, we will try to integrate interactions that are more realistic, and to initialise the population of agents with the help of the results of the ”voter autrement” experimentation \cite{vot2017}.

\section{OUR MODEL} \label{sec:mod}

  We choose to model each voter as a point in a 11 dimensional space, and we initialise the model with the study’s continuous evaluation results, giving as coordinates to a voter the marks he has given to the different candidates. Because we also know the official vote of each voter, we integrate the candidates into that space by taking the centroids of their voter’s continuous evaluation. The model is dynamic and at each iteration, the voters move and then a plurality vote is computed. To compute this vote, we make each voter vote for the closest candidate from him and the candidate with the most votes is the winner. In order to make the agents interact between each other, we use polls that give information to everyone on the current plurality vote outcome. The voter compute their move taking into account the results of the last poll. That’s the way the agents interact between each other.

The model ran with different set of utility functions and agent behaviour. Basically, we make each agent move with a velocity vector computed at each iteration, and several methods were used to compute this one. We will discuss those methods and their expected properties in this section.

\enlargethispage{30pt}
\subsection{Formal notations}
  Formally, we note as $V_i$ the set of voters at iteration $i$, and $C$ the (static) set of candidates. Both of these sets contain a finite number of 11-dimension vectors. Thus $v_{k,i}$ is the position of voter $v_k$ at iteration $i$. $s_i$ is the last poll available at iteration $i$, also represented as an 11-dimension vector, containing the number of votes for each candidate in the poll’s sample.

$V(v_{k,i} , s_i)$ is the velocity vector of voter $v$, knowing its current position and the result of the last poll. At each iteration, the next position vector of each voter $v_{k,i+1} \in V_{i+1}$ is given by the equation (1), as the sum of its previous position and its velocity vector computed from this position and the last poll.
\begin{equation}
    v_{k,i+1} = v_{k,i} + V(v_{k,i}, s_i)
\end{equation}

\subsection{Utility functions}

Before introducing the movement rules of our model, we introduce the concept of utility function that we will later use in these rules. A utility function is defined for each voter, as the utility a candidate would have for this voter if he was elected. 
We define a utility function $u: \mathbb{R}^{11} \times \mathbb{R}^{11} \rightarrow \mathbb{R}$, giving for the first point (a voter) the utility that the second point (a candidate) would have if winning the election. In practice, we will use $u(v_{k,i}, c)$ for $v_{k,i} \in V_i$ and $c \in C$.

In our model, we want that utility function to be in line with the fact that the voters vote for the closest candidate from them. So intuitively, the closer a voter is to a candidate, the more he likes him.
The main property that we want from these functions is that they are decreasing functions of the distance between the voter and the candidates, and that they are not linear (to model that some voters are strongly attached to a candidate). 

Noting $dist(v,c)$ the Euclidean distance between voter $v$ and candidate $c$, we designed 2 different utility functions representing different reasoning. The first utility function given by equation (2) below simply expresses utility as inversely proportional to the distance with the candidate.
\begin{equation}
   u: v,c \rightarrow \frac{1}{dist(v,c)}
\end{equation}

The second utility function that we designed also models the fact that a voter can be strongly opposed to a candidate when he is too far away from him (further than a parameter $\alpha$). In this case this utility function becomes negative, ensuring that the voter will not get any closer to this candidate (repulsion). This second function is given by equation (3) below. The drawback is that some voters just flee away from the whole set of candidates, which we model as abstention in the vote.
\begin{equation}
    u: v,c \rightarrow \frac{\alpha - dist(v,c)}{(1 + dist(v,c))^2}
\end{equation}

 \enlargethispage{20pt}
\subsection{Velocity computation}
  
In this section we describe three techniques that can be used to compute the velocity vector of the voters.

\paragraph{First approach: 3-pragmatist rule}
A first approach is to use a rule described in \cite{agp17} which is the three pragmatist rule. This rule makes the voter vote for his favourite candidate amongst the first three candidates of the election, which represent a strategic vote for a candidate who has a chance to win. We integrate that in our model by making each voter move towards this candidate, once given the results of the last poll. The 3-pragmatist velocity vector towards candidate $c$, the closest to voter $v_{k,i}$ among the first 3 candidates in results of poll $s_i$, is given by the following equation:
\begin{equation}
    V(v_{k,i}, s_i) = e_{v_{k,i},c}
\end{equation}
where $e_{a,b}$ is the unit vector from point $a$ to $b$.

\paragraph{Second approach: maximum expected utility}
A second approach is to use a utility function and to move toward the candidate that maximises the expected utility. In order to approximate this expectation, we use the poll results to compute a probability that each candidate wins, given by the following equation as the proportion of expressed votes in favour of this candidate.
\begin{equation}
    p(wins(c) \vert s_i) = \frac{number of votes for c in s_i}{total number of votes in s_i}
\end{equation}

Once we have the probability that each candidate wins, computed from the polls, the expected utility of their victory is the product of this probability by the utility of them winning. This approximation gives a chance even to candidates who have a small number of votes, provided that the utility of them winning is high enough to compensate. On the contrary, other approaches would just dismiss these candidates altogether, considering that voting for them is useless because they have no chance to win in the end. 

Thus we can compute the candidate selected by the voter as the one that maximises this expected utility, as expressed by the following equation:
\begin{equation}
    c = argmax_{c \in C} [u(v_{k,i}, c) * p(wins(c) \vert s_i)]
\end{equation}

In the end the velocity vector is computed as the unit vector from voter $v_{k,i}$ to that candidate $c$:
\begin{equation}
    V(v_{k,i}, s_i) = e_{v_{k,i},c}
\end{equation}

\paragraph{Third approach: personalised opinion centre}
The third approach is different in the sense that rather than moving towards a single candidate, it makes the voter move towards a personalised opinion centre. In this approach we compute velocity
as a linear combination of the expected utilities of the different candidates, as expressed by the following equation.
\begin{equation}
    V(v_{k,i}, s_i) = \sum_{c \in C} e_{v_{k,i},c} * u(v_{k,i}, c) * p(wins(c) \vert s_i)
\end{equation}

Since movement towards each candidate is weighed by the expected utility of that candidate, we can expect that this vector will somehow be pointing in a direction that will make the voter closer to the candidate that maximises its expected utility.

\section{RESULTS} \label{sec:rez}

\subsection{Visualization}

\begin{figure}[hbt]
    \centering
    \includegraphics[scale=0.4]{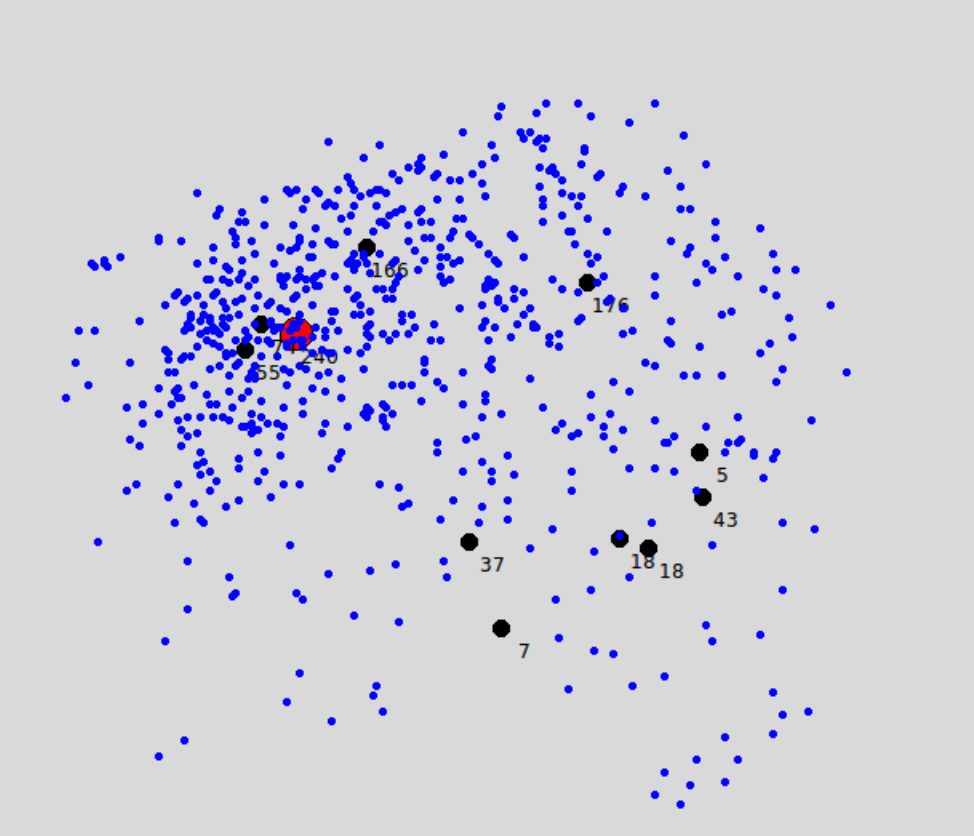}
    \caption{Principal Component Analysis to initialise the model}
    \label{fig:fig0}
\end{figure}

In order to visualise the multi-agent model behaviour, we choose to use dimensionality reduction techniques. Two of them were tested, PCA (Principal Component Analysis, \cite{pca}) and T-SNE (t-distributed stochastic neighbor embedding, \cite{tsne}). We choose PCA because it is less computationally expensive and is not stochastic. The PCA results on initialisation of the model are shown on Figure~\ref{fig:fig0}.

\subsection{Parameters}
Before presenting the results, we will review the set of parameters used and the circumstances of the simulations. The simulations were ran on a sample of 1000 voters which are drawn among the 30000 voters who took part in the ``Voter Autrement'' experiment. We will use the same seed for the random generator for this sample to allow reproducibility of the experiments. We have defined above two utility functions and three velocity formulas that can be tested. We are not going to present nor simulate all the combinations, but we will rather focus on the most interesting ones, and try to identify the impact of the parameters on the simulation. 

Below are the official votes given by the sample’s voters. Compared to the actual results of the 2017 presidential election, it is clearly not representative, but it will still be interesting to compare those results to the outputs of our model.
\small
\begin{itemize}
    \item Jean-Luc Melenchon 403 
    \item Jean Lassalle 7
    \item François Asselineau 5
    \item Nicolas Dupont-Aignan 11
    \item François Fillon 34
    \item Benoit Hamon 144 
    \item Philippe Poutou 12
    \item Emmanuel Macron 200 
    \item Marine Le Pen 18
    \item Nathalie Arthaud 3 
    \item Jacques Cheminade 2
\end{itemize}
\normalsize

\subsection{Three-pragmatist rule}

We tried the three-pragmatist velocity rule with different poll results. We can see that this rule is very sensitive to the order of candidates given by the poll (because it only draws electors towards the first three candidates in the results of the poll). 

\begin{figure}[hbt]
    \centering
    \includegraphics[scale=0.3]{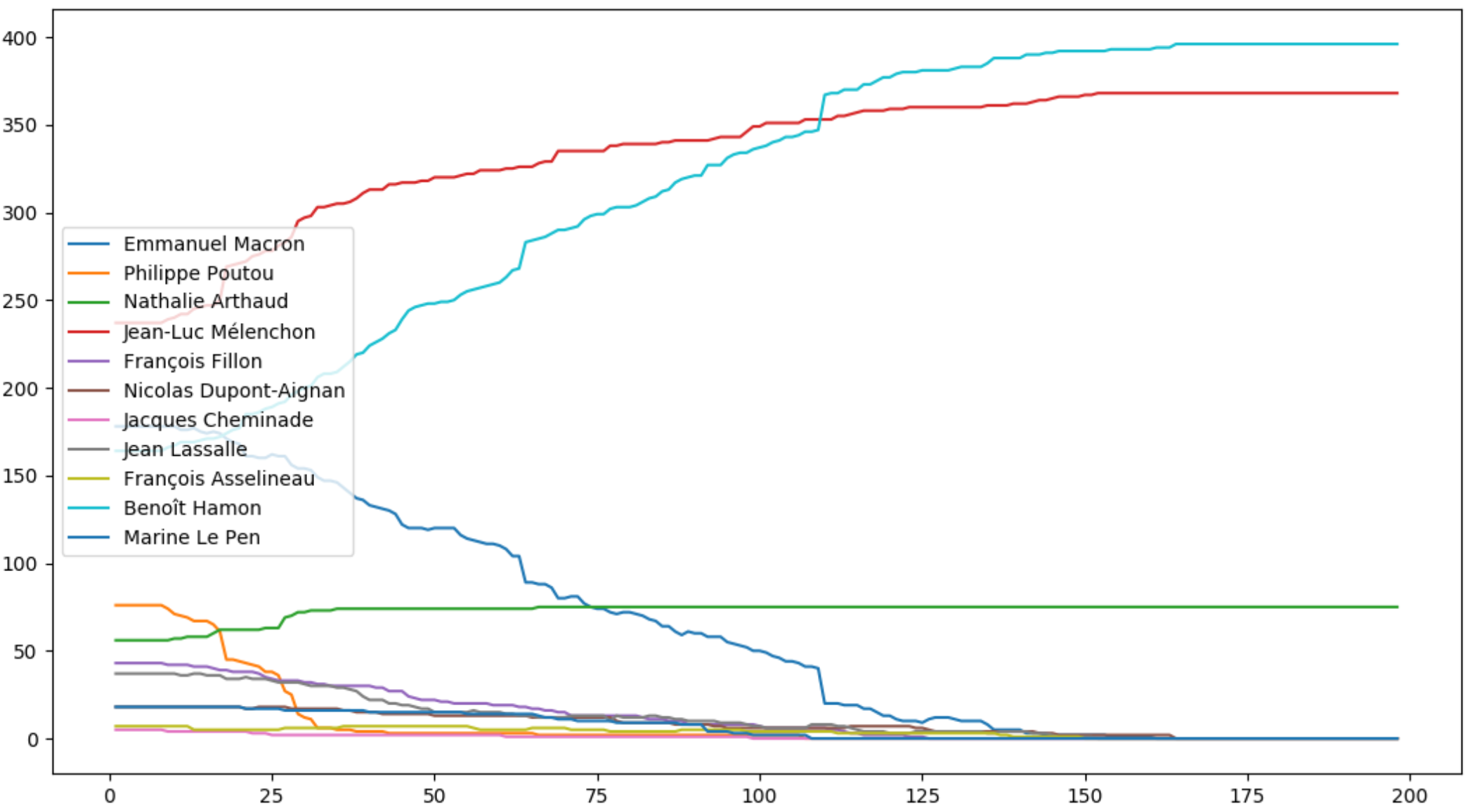}
    \caption{With a poll where Nathalie Arthaud is above Emmanuel Macron}
    \label{fig:fig1}
\end{figure}

Indeed, with a poll giving 
Nathalie Arthaud above Emmanuel Macron (Figure~\ref{fig:fig1}, the voters closer to Macron (who is not in the top 3) move toward their closest candidate among the three winners, in this case Benoit Hamon. This leads to Hamon winning the election against Jean-Luc Melenchon. 

\begin{figure}[hbt]
    \centering
    \includegraphics[scale=0.3]{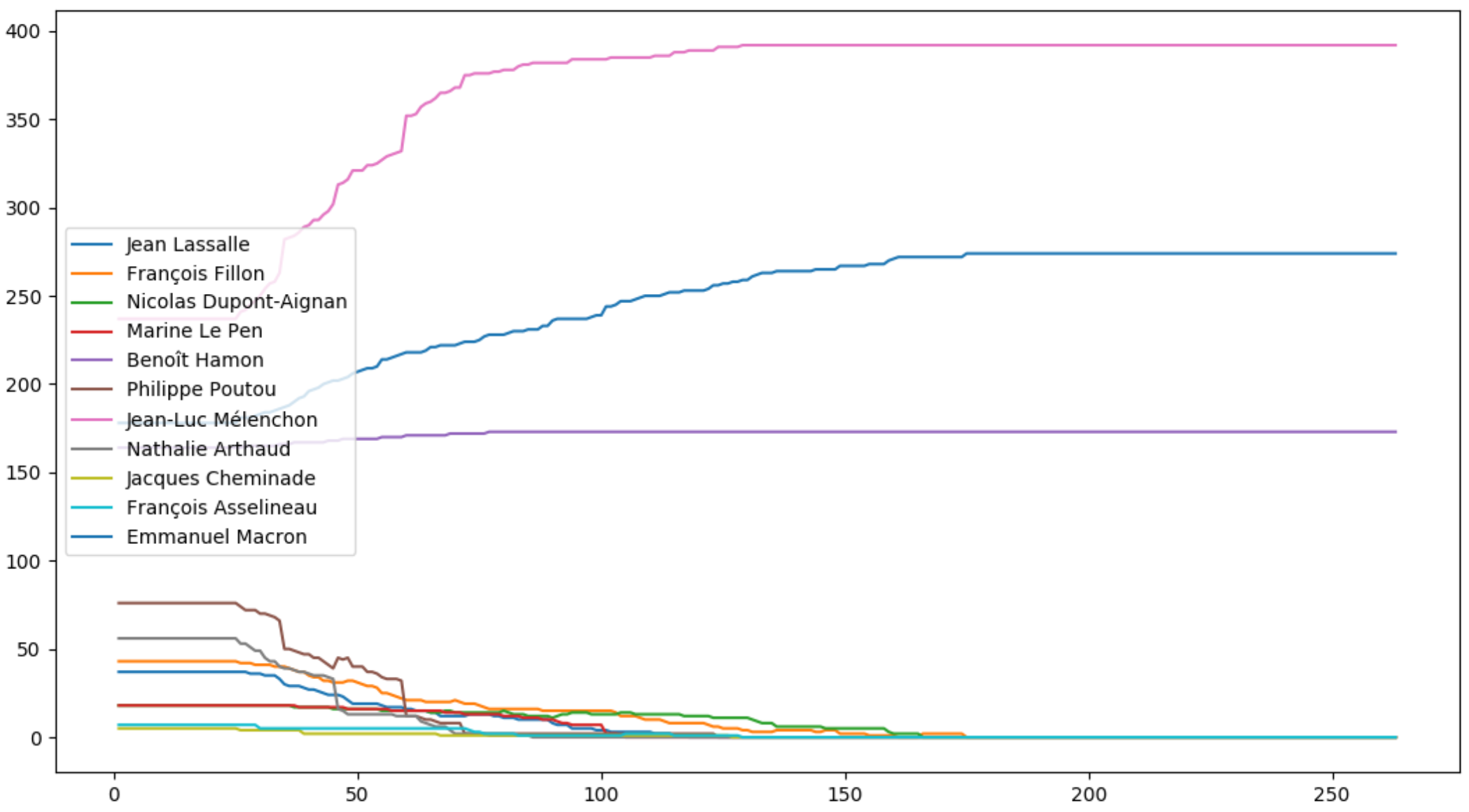}
    \caption{With a poll where Emmanuel Macron is above Nathalie Arthaud}
    \label{fig:fig2}
\end{figure}

On the contrary, if the order in the poll results is the actual order (Macron above Arthaud), the real order is just confirmed by the next iterations (Figure~\ref{fig:fig2}).

This rule gives accurate results regarding the actual official votes of our sample, but only for the first three candidates of the poll. It also gives totally wrong results if the poll order is not the same as the actual order. In fact, some people behave in such a way that they will not change their vote, even if their candidate has no chance to win at all. Explanations can be a desire to show contestation, or a strong attachment to this candidate and their ideas.

\subsection{Maximum expected utility}

\begin{figure}[hbt]
    \centering
    \includegraphics[scale=0.3]{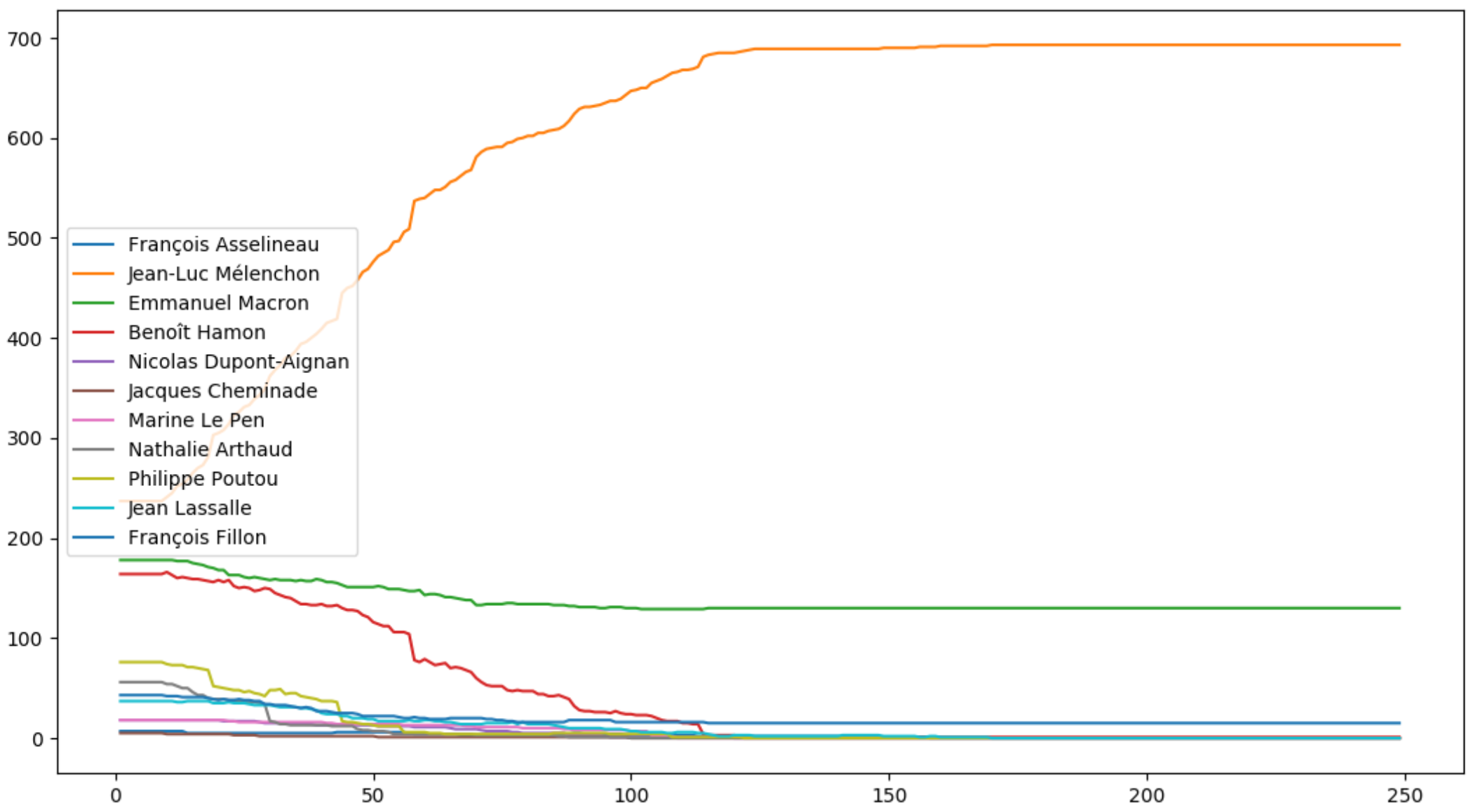}
    \caption{with utility function (2) and one poll}
    \label{fig:fig3}
\end{figure}

In our next experiment, we used our second expected utility function and one poll, and computed velocity with the second technique, of maximising expected utility. Results are shown on Figure~\ref{fig:fig3}. 

In this experiment, Emmanuel Macron has nearly as many votes as expected in the sample. However, the utility function still needs to be tuned, because all the voters who initially preferred Benoit Hamon, reported their votes (after the poll) towards Jean-Luc Melenchon, which is not the case in the real sample results. One explanation is that our expected utility model is imperfect; indeed, people have a non-objective way of thinking in this situation, and this is hard to model in a homogeneous way.

\subsection{Utility speed (c)}

In the next scenario, we used the third velocity formula, that moves voters towards a personalised opinion centre. We compared the results obtained with our 2 utility functions, and also with a single poll or regular polls along the campaign.

\begin{figure}[hbt]
    \centering
    \includegraphics[scale=0.3]{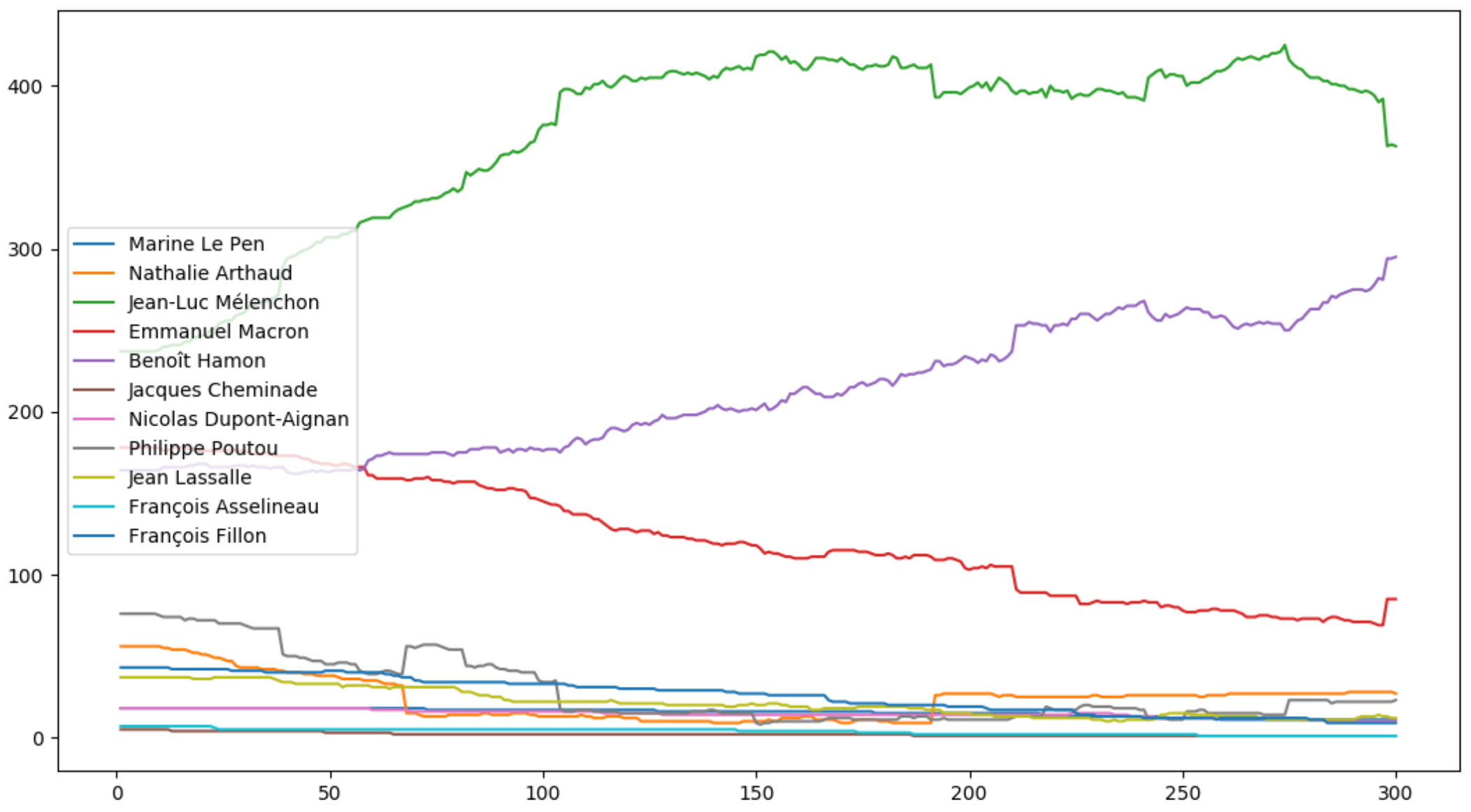}
    \caption{With utility function (1) and one poll}
    \label{fig:fig4}
\end{figure}

Figure~\ref{fig:fig4} illustrates the results with utility directly inversely proportional to distance (first formula), and one single poll. In this figure, the number of votes for Jean-Luc Melenchon is accurate, but Emmanuel Macron's votes are too low.

\begin{figure}[hbt]
    \centering
    \includegraphics[scale=0.3]{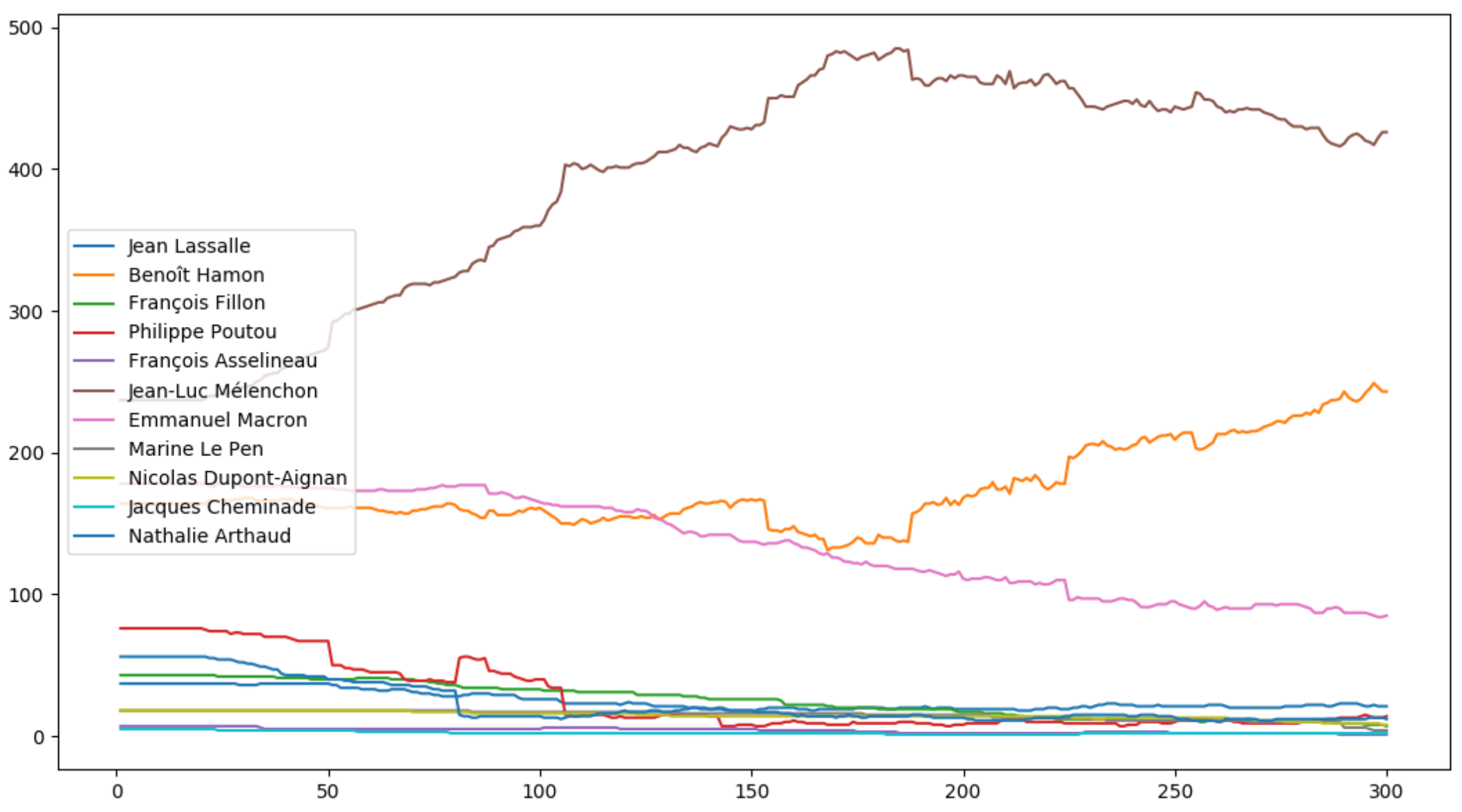}
    \caption{with utility function (1) and one poll every 20 iteration}
    \label{fig:fig5}
\end{figure}

With that same utility function, regular polls do not greatly change the results (Figure~\ref{fig:fig5}). We can observe nearly the same behaviour as in the previous figure, even though we updated the information available to the agents every 20 iterations.

\begin{figure}[hbt]
    \centering
    \includegraphics[scale=0.3]{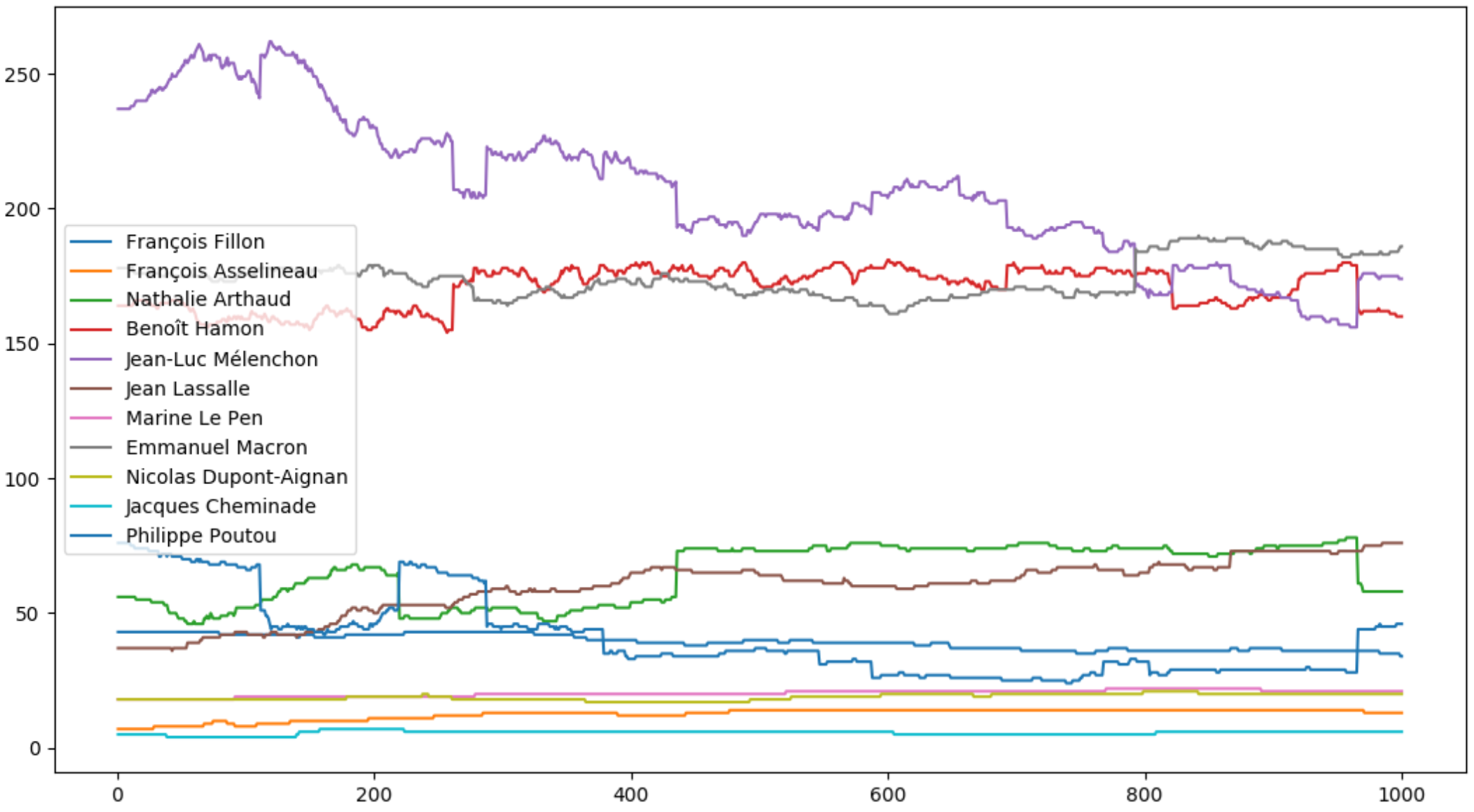}
    \caption{With utility function (2) and one poll every 20 iteration}
    \label{fig:fig6}
\end{figure}

\newpage
Finally we also tested the second utility function introducing repulsion towards candidates that are too far, still with regular update polls. In Figure~\ref{fig:fig6} the results are much more tied, but they are also less accurate with respect to the actual official votes of the sample. It is also hard to conclude on a final order in this case, given that the rule does not converge.

\subsection{Periodicity of polls}

The simulations run show that periodical polls do not fundamentally change the result of the election. Indeed, quite often the results of the polls (here made on 10\% of the total population) are consistent with the order of the real results, so the movement tendency and the order of candidates are just confirmed a bit more by each new poll. And whenever a poll gives a different order, it is often counter-balanced by the next poll. 

However, in a real situation, other events can have an effect on the voters preferences, such as media scandals (for instance the Penelope gate scandal involving François Fillon during the French 2017 presidential elections, which probably made him lose that election). In this case, even if some people are not directly deterred from voting for their favourite candidate by the scandal, they may be indirectly led to do so by strategy, if the subsequent polls show that this candidate's chance to win has dropped too much.

\subsection{Notes about properties} 

The movement rules that we presented above were mainly built to make the voters in the simulation behave somehow close to how the actual voters behave. Because we focused on those properties, and did not make any mathematical analysis of the rules, they do not necessarily have good mathematical properties, in particular convergence.

It is hard to interpret the results if the simulations are not converging towards at least a fixed order of the candidates, if not a fixed number of votes. Indeed, we have no way to link the number of iterations in the simulation to a real duration (in days), so any interpretation would remain a strong supposition. The results presented here have been obtained after a large number of iterations, when the electors are ”caught in a pit” and mainly move towards one candidate, so that the order of the candidates does converge, but the convergence of the number of votes is not guaranteed.

\section{CONCLUSION AND OPENING} \label{sec:cci}

The models presented above have shown different interesting behaviours. The one which gave the most accurate results is the simplest one, using the three-pragmatist rule, even though it does not allow more than 3 candidates to receive votes. To extend this approach it would be interesting to experiment a k-pragmatist rule for different values of k, and play around it (maybe compute a velocity vector using different k-pragmatist results). 

The maximum expected utility rule should have really good properties, but the utility functions were chosen quite arbitrarily. We struggled to accurately compute an expected utility, because people do not really conduce complex probabilistic calculus, and it is hard to model what really happens in someones’ mind when making a voting decision. Many factors might play a role such as emotions or cognitive biases.
Also, the second utility function uses a customisable $\alpha$ parameter to represent the tolerance before a voter is repulsed by a candidate. Here we used the same $\alpha$ for each voter, but we could also compute an individual value for each voter, as a function of his starting position. 

Finally, in this paper we only gave information to the voters through polls, unique or recurrent, but more complex events could be modelled, as for example media coverage of scandals, the use of social networks to communicate, or the influence of other voters on each voter's opinions. 




\end{document}